%% file: sample-authordraft.tex
\documentclass[sigconf]{acmart}

\AtBeginDocument{%
  \providecommand\BibTeX{{%
    \normalfont B\kern-0.5em{\scshape i\kern-0.25em b}\kern-0.8em\TeX}}}

\setcopyright{acmcopyright}
\copyrightyear{2021}
\acmYear{2021}
\acmDOI{10.1145/1122445.1122456}

\acmConference[Woodstock '18]{Woodstock '18: ACM Symposium on Neural Gaze Detection}{June 03--05, 2018}{Woodstock, NY}
\acmBooktitle{Woodstock '18: ACM Symposium on Neural Gaze Detection, June 03--05, 2018, Woodstock, NY}
\acmPrice{15.00}
\acmISBN{978-1-4503-XXXX-X/18/06}

\usepackage{romannum}
\usepackage{textcomp}
\usepackage{tabularx}
\usepackage{diagbox}

\usepackage{bbding}
\usepackage{pifont}
\usepackage{wasysym}

\usepackage{listings}
\usepackage{color}
\definecolor{dkgreen}{rgb}{0,0.6,0}
\definecolor{gray}{rgb}{0.5,0.5,0.5}
\definecolor{mauve}{rgb}{0.58,0,0.82}
\definecolor{backcolour}{rgb}{0.95,0.95,0.92}
\usepackage{graphics}
\usepackage{caption}
\usepackage{subcaption}
\graphicspath{{./figures/}}

\begin{document}

\pagenumbering{arabic}
\pagestyle{plain}
\settopmatter{printfolios=true}


\newcommand{\ada}[1]{\textcolor{red}{AG: #1}}
\newcommand{\ketan}[1]{\textcolor{blue}{KB: #1}}
\newcommand{\jin}[1]{\textcolor{green}{JH: #1}}

\title{Poster: Enabling Flexible Edge-assisted XR}

\author{Jin Heo}
\affiliation{%
  \institution{Georgia Institute of Technology}
  \city{Atlanta}
  \state{Georgia}
  \country{USA}}
\email{jheo33@gatech.edu}

\author{Ketan Bhardwaj}
\affiliation{%
  \institution{Georgia Institute of Technology}
  \city{Atlanta}
  \state{Georgia}
  \country{USA}}
\email{ketanbj@gatech.edu}

\author{Ada Gavrilovska}
\affiliation{%
  \institution{Georgia Institute of Technology}
  \city{Atlanta}
  \state{Georgia}
  \country{USA}}
\email{ada@cc.gatech.edu}

\renewcommand{\shortauthors}{Jin, et al.}

\begin{abstract}
Extended reality (XR) is touted as the next frontier of the digital future. XR includes all immersive technologies of augmented reality (AR), virtual reality (VR), and mixed reality (MR).
XR applications obtain the real-world context of the user from an underlying system,  and provide rich, immersive, and interactive virtual experiences based on the user's context in real-time.
XR systems process streams of data from device sensors, and provide functionalities including perceptions and graphics required by the applications.
These processing steps are computationally intensive, and the challenge is that they must be performed within the strict latency requirements of XR.
This poses limitations on the possible XR experiences that can be supported on mobile devices with limited computing resources.

In this XR context, edge computing is an effective approach to address this problem for mobile users.
The edge is located closer to the end users and enables processing and storing data near them.
In addition, the development of high bandwidth and low latency network technologies such as 5G facilitates the application of edge computing for latency-critical use cases \cite{chen2017empirical, hu2016quantifying}.
This work presents an XR system for enabling flexible edge-assisted XR.
\end{abstract}


\settopmatter{printacmref=false}



\maketitle

\input{sections/sections.tex}

\bibliographystyle{ACM-Reference-Format}
\bibliography{sample-base}

\end{document}

%% file: sections/sections.tex
\section{Introduction}
There have been efforts for XR systems to utilize remote servers for their workload \cite{lai2019furion, meng2020coterie, liu2020firefly, liu2018cutting, zhang2019rendering, liu2019edge, george2020openrtist, schneider2017augmented, cloudxr, azremoterendering, isar, azcustomvision}.
A common thread across these systems is that they make a priori decisions about the expensive functionalities whose offload should be enabled, based on assumptions about operating or environmental factors such as the client device capabilities and network conditions.
In that sense, the resulting distributed architectures provide support for offloading of concrete functional components, and their benefits can be realized only in specific deployment contexts.
We argue, however, that as XR gains in popularity, the target deployment contexts will differ vastly, in terms of the end device and edge server processing capabilities, the connectivity among them, the functional elements of the XR stacks, and the specific components that comprise concrete applications.
There has been system research for flexible computation offloading \cite{cuervo2010maui, chun2011clonecloud}, and they enable seamlessly running the functions on the server of the same environment as the client via the function call interface such as RPC.
However, XR workloads involve multimedia data processing in multiple steps, and the RPC-based executions are hardly efficient.
Moreover, instead of seamless function offloading, we want the server to fully utilize their available heterogeneous resources.

To alleviate this issue, we propose a flexibly configurable and high-performance system for edge-assisted XR that enables tailored and effective edge assistance for each user.
The fundamental goal is, therefore, to make XR system tasks offlodable and users optimally utilizing additional computing power at remote servers for better user experiences.
Beyond simply offloading and running the codes of tasks on the server, the tasks should be able to benefit from the resources available on the edge such as hardware accelerators.
While providing such flexibility, our system needs to be designed highly efficiently for the performance requirements of XR use cases.

\section{Initial Design}
Motivated by the success of stream processing approaches, we are building our system by applying the stream parallel processing paradigm.
Stream processing has advantages to flexibility, heterogeneous resource utilization, and high throughput in that it
requires each functionality implemented as an independent unit, a compute kernel, and enables pipeline parallelism for the kernel execution \cite{roger2019combining, beard2017raftlib}.
XR functionalities are implemented as a set of compute kernels in a modular manner, and they are pipelined with data communication ports.
Our system kernel and port interfaces are designed to make each kernel an offloadable unit, whose ports can be dynamically configured for remote or local at deployment time.
Thus, it can be viewed as a high-performance distributed stream processing system (DSPS) for latency-sensitive and real-time use cases including XR with respect to its system design.

\begin{figure}[h]
  \centering
  \includegraphics[width=\linewidth]{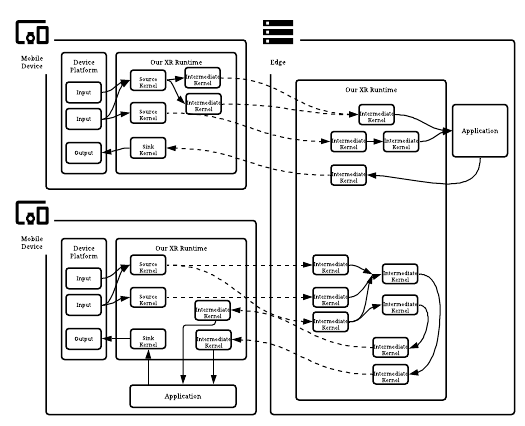}
  \caption{The high-level illustration of the system architecture.}
  \label{fig:flexr}
\end{figure}

Figure \ref{fig:flexr} shows the example architecture of our system.
Developers implement and register their kernels in advance, and these available kernels are flexibly distributed at runtime by the deployers.
Deployers can activate each kernel's connections for local and remote for their distributed pipeline; in Figure \ref{fig:flexr}, the remote port connections are represented as dotted lines.
Among the kernels, device source and sink kernels have to be placed on the client device while the intermediate kernels can be distributed.

\section{Preliminary Results}
Even though our system, as a DSPS, gets benefitted from the design of stream processing for pipeline offloading and parallelism, this design can increase the end-to-end processing latency because of the communication costs between kernels \cite{beard2017raftlib, khare2019linearize, basanta2017patterns}.
We must address this issue to make our system highly efficient for XR use cases of strict latency requirements.
Our system minimizes the local communication costs of connected kernels through zero-copy communication.
In addition, it provides efficient compression functionalities and communication protocols for multimedia data to minimize remote communication costs across the entire distributed pipeline.

There are widely used DSP engines: Apache Storm \cite{toshniwal2014storm}, Apache Flink \cite{carbone2015apache}, Apache Spark Streaming \cite{zaharia2012discretized}.
While these DSP frameworks are successful for application domains such as big data and the Internet of Things, these frameworks have been specialized for high throughput rather than latency by using techniques of various batching streaming models, as previous studies showed \cite{lopez2016performance, karimov2018benchmarking, lu2014stream, karakaya2017comparison}.
Instead, we develop our system by extending the local stream processing library, RaftLib \cite{beard2017raftlib}.
For the local stream processing libraries, we explore GStreamer \cite{gstreamer} and RaftLib \cite{beard2017raftlib}.
Along with these two stream processing libraries, we also consider the frameworks which are widely used by other research communities: Python pipeline libraries \cite{mpipe, pypeln} and the robot operating system (ROS) \cite{quigley2009ros}.

\begin{table}[]
  \caption{\label{tab:localcommunication} The local communication latencies averaged over 1000 frame transmissions in milliseconds.}
  \begin{tabular}{|l|c|c|c|c|}
    \hline
    \diagbox{Libs}{Resolution} & 720p & 1080p & 1440p & 2160p \\ \hline
    ROS Pub/Sub\cite{quigley2009ros}        & 3.4           & 6.9            & 7.1            & 12.5           \\ \hline
    ROS Shm Pub/Sub \cite{rosshm}    & 2.2           & 4.3            & 5.9            & 10.2           \\ \hline
    Python Queue \cite{mpipe, pypeln}      & 14.3          & 24.1           & 30.4           & 52.1           \\ \hline
    Python Pipe \cite{mpipe, pypeln}        & 9.3           & 17.1           & 29.5           & 52.1           \\ \hline
    Python Shm (PyObj) & 0.1(3.0)      & 1.0(8.6)       & 1.4(14.8)      & 2.8(32.3)      \\ \hline
    GStreamer \cite{gstreamer}         & 0.1           & 0.1            & 0.1            & 0.1            \\ \hline
    Ours with RaftLib \cite{beard2017raftlib}  & 0.1           & 0.1            & 0.1            & 0.1            \\ \hline
  \end{tabular}
\end{table}

We perform experiments measuring the local data transmission latencies for the aforementioned frameworks with a pipeline of two kernels and one data link transmitting raw RGB frames of four different resolutions.
Table \ref{tab:localcommunication} shows the averaged latencies of 1000 transmissions of different resolution images on each runtime.
ROS and Python provide the process-level pipelining with IPC and show the high latencies for data transmissions even with shared-memory IPC.
For Python SharedMemory (available from version 3.8), the data is visible as a byte-array object to the receiving process at lower latencies.
However, when the byte-array object is converted to a frame-type object in the receiving process, it results in huge latencies for copying the data from the shared memory and creating a new PyObject shown as `Python Shm (PyObj)'.
In ROS, the basic ROS message communication is based on the publish-subscribe model via socket interfaces.
Although the performance of socket-based IPC was much better than the Python communication channels, this socket-based IPC also requires data copying and shows high overhead with high-resolution frames.
Additionally, we check the ROS extension package for shared memory IPC \cite{rosshm}, `ROS Pub\slash Sub Shm', but it
doesn't totally remove the overhead; it has an additional channel for synchronizing the access to the shared memory from publishers and subscribers, and also copies data to the shared memory for publishing.

GStreamer and ours provide the thread-level pipelining with zero-copy-based communication methods.
Although GStreamer can be used for efficient SPS runtime, it is challenging to extend GStreamer into a DSPS for XR.
GStreamer, as a multimedia framework, requires additional kernels for multi-stream pipe-fitting, synchronization, and remote communication in Table \ref{tab:gstdesdiff} due to its strict requirement for stream synchronization \cite{taymans2013gstreamer}.
Our system's kernels can be used for setting up distributed XR pipeline architectures without auxiliary kernels via relaxed synchronization and finer communication port configuration mechanisms.

\begin{table}[]
  \caption{\label{tab:gstdesdiff} The design feature differences between our system and GStreamer for distributed XR pipelines.}
  \begin{tabular}{|l|c|c|}
    \hline
    & Ours & GStreamer \\ \hline
    \begin{tabular}[c]{@{}l@{}}Not requiring additional kernels for\\ multi input/output streams\end{tabular}   &   \ding{52}   &          \\ \hline
    \begin{tabular}[c]{@{}l@{}}Not requiring additional kernels\\ for sync/async stream inputs \end{tabular}    &   \ding{52}   &          \\ \hline
    \begin{tabular}[c]{@{}l@{}}Not requiring additional kernels\\ for remote communication \end{tabular}        &   \ding{52}   &          \\ \hline
  \end{tabular}
\end{table}

\section{Summary and Next Steps}
We describe our motivation for this work and present preliminary results for supporting our system design for flexible edge assistance.
However, we need to show the practicality of our system with an end-to-end XR use case.

%% file: sample-authordraft.bbl

\begin{thebibliography}{33}


\ifx \showCODEN    \undefined \def \showCODEN     #1{\unskip}     \fi
\ifx \showDOI      \undefined \def \showDOI       #1{#1}\fi
\ifx \showISBNx    \undefined \def \showISBNx     #1{\unskip}     \fi
\ifx \showISBNxiii \undefined \def \showISBNxiii  #1{\unskip}     \fi
\ifx \showISSN     \undefined \def \showISSN      #1{\unskip}     \fi
\ifx \showLCCN     \undefined \def \showLCCN      #1{\unskip}     \fi
\ifx \shownote     \undefined \def \shownote      #1{#1}          \fi
\ifx \showarticletitle \undefined \def \showarticletitle #1{#1}   \fi
\ifx \showURL      \undefined \def \showURL       {\relax}        \fi
\providecommand\bibfield[2]{#2}
\providecommand\bibinfo[2]{#2}
\providecommand\natexlab[1]{#1}
\providecommand\showeprint[2][]{arXiv:#2}

\bibitem[\protect\citeauthoryear{Basanta-Val, Fernandez-Garcia,
  Sanchez-Fernandez, and Arias-Fisteus}{Basanta-Val et~al\mbox{.}}{2017}]%
        {basanta2017patterns}
\bibfield{author}{\bibinfo{person}{Pablo Basanta-Val},
  \bibinfo{person}{Norberto Fernandez-Garcia}, \bibinfo{person}{Luis
  Sanchez-Fernandez}, {and} \bibinfo{person}{Jesus Arias-Fisteus}.}
  \bibinfo{year}{2017}\natexlab{}.
\newblock \showarticletitle{Patterns for distributed real-time stream
  processing}.
\newblock \bibinfo{journal}{\emph{IEEE Transactions on Parallel and Distributed
  Systems}} \bibinfo{volume}{28}, \bibinfo{number}{11} (\bibinfo{year}{2017}),
  \bibinfo{pages}{3243--3257}.
\newblock


\bibitem[\protect\citeauthoryear{Beard, Li, and Chamberlain}{Beard
  et~al\mbox{.}}{2017}]%
        {beard2017raftlib}
\bibfield{author}{\bibinfo{person}{Jonathan~C Beard}, \bibinfo{person}{Peng
  Li}, {and} \bibinfo{person}{Roger~D Chamberlain}.}
  \bibinfo{year}{2017}\natexlab{}.
\newblock \showarticletitle{RaftLib: a C++ template library for high
  performance stream parallel processing}.
\newblock \bibinfo{journal}{\emph{The International Journal of High Performance
  Computing Applications}} \bibinfo{volume}{31}, \bibinfo{number}{5}
  (\bibinfo{year}{2017}), \bibinfo{pages}{391--404}.
\newblock


\bibitem[\protect\citeauthoryear{Carbone, Katsifodimos, Ewen, Markl, Haridi,
  and Tzoumas}{Carbone et~al\mbox{.}}{2015}]%
        {carbone2015apache}
\bibfield{author}{\bibinfo{person}{Paris Carbone}, \bibinfo{person}{Asterios
  Katsifodimos}, \bibinfo{person}{Stephan Ewen}, \bibinfo{person}{Volker
  Markl}, \bibinfo{person}{Seif Haridi}, {and} \bibinfo{person}{Kostas
  Tzoumas}.} \bibinfo{year}{2015}\natexlab{}.
\newblock \showarticletitle{Apache flink: Stream and batch processing in a
  single engine}.
\newblock \bibinfo{journal}{\emph{Bulletin of the IEEE Computer Society
  Technical Committee on Data Engineering}} \bibinfo{volume}{36},
  \bibinfo{number}{4} (\bibinfo{year}{2015}).
\newblock


\bibitem[\protect\citeauthoryear{Chen, Hu, Wang, Zhao, Amos, Wu, Ha, Elgazzar,
  Pillai, Klatzky, et~al\mbox{.}}{Chen et~al\mbox{.}}{2017}]%
        {chen2017empirical}
\bibfield{author}{\bibinfo{person}{Zhuo Chen}, \bibinfo{person}{Wenlu Hu},
  \bibinfo{person}{Junjue Wang}, \bibinfo{person}{Siyan Zhao},
  \bibinfo{person}{Brandon Amos}, \bibinfo{person}{Guanhang Wu},
  \bibinfo{person}{Kiryong Ha}, \bibinfo{person}{Khalid Elgazzar},
  \bibinfo{person}{Padmanabhan Pillai}, \bibinfo{person}{Roberta Klatzky},
  {et~al\mbox{.}}} \bibinfo{year}{2017}\natexlab{}.
\newblock \showarticletitle{An empirical study of latency in an emerging class
  of edge computing applications for wearable cognitive assistance}. In
  \bibinfo{booktitle}{\emph{Proceedings of the Second ACM/IEEE Symposium on
  Edge Computing}}. \bibinfo{pages}{1--14}.
\newblock


\bibitem[\protect\citeauthoryear{Chun, Ihm, Maniatis, Naik, and Patti}{Chun
  et~al\mbox{.}}{2011}]%
        {chun2011clonecloud}
\bibfield{author}{\bibinfo{person}{Byung-Gon Chun}, \bibinfo{person}{Sunghwan
  Ihm}, \bibinfo{person}{Petros Maniatis}, \bibinfo{person}{Mayur Naik}, {and}
  \bibinfo{person}{Ashwin Patti}.} \bibinfo{year}{2011}\natexlab{}.
\newblock \showarticletitle{Clonecloud: elastic execution between mobile device
  and cloud}. In \bibinfo{booktitle}{\emph{Proceedings of the sixth conference
  on Computer systems}}. \bibinfo{pages}{301--314}.
\newblock


\bibitem[\protect\citeauthoryear{{Cristian Garcia}}{{Cristian Garcia}}{2018}]%
        {pypeln}
\bibfield{author}{\bibinfo{person}{{Cristian Garcia}}.}
  \bibinfo{year}{2018}\natexlab{}.
\newblock \bibinfo{title}{Pypeln, A simple yet powerful Python library for
  creating concurrent data pipelines}.
\newblock \bibinfo{howpublished}{\url{https://cgarciae.github.io/pypeln/}}.
\newblock
\newblock
\shownote{[Online; accessed 13-June-2021].}


\bibitem[\protect\citeauthoryear{Cuervo, Balasubramanian, Cho, Wolman, Saroiu,
  Chandra, and Bahl}{Cuervo et~al\mbox{.}}{2010}]%
        {cuervo2010maui}
\bibfield{author}{\bibinfo{person}{Eduardo Cuervo}, \bibinfo{person}{Aruna
  Balasubramanian}, \bibinfo{person}{Dae-ki Cho}, \bibinfo{person}{Alec
  Wolman}, \bibinfo{person}{Stefan Saroiu}, \bibinfo{person}{Ranveer Chandra},
  {and} \bibinfo{person}{Paramvir Bahl}.} \bibinfo{year}{2010}\natexlab{}.
\newblock \showarticletitle{Maui: making smartphones last longer with code
  offload}. In \bibinfo{booktitle}{\emph{Proceedings of the 8th international
  conference on Mobile systems, applications, and services}}.
  \bibinfo{pages}{49--62}.
\newblock


\bibitem[\protect\citeauthoryear{George, Eiszler, Iyengar, Turki, Feng, Wang,
  Pillai, and Satyanarayanan}{George et~al\mbox{.}}{2020}]%
        {george2020openrtist}
\bibfield{author}{\bibinfo{person}{Shilpa George}, \bibinfo{person}{Thomas
  Eiszler}, \bibinfo{person}{Roger Iyengar}, \bibinfo{person}{Haithem Turki},
  \bibinfo{person}{Ziqiang Feng}, \bibinfo{person}{Junjue Wang},
  \bibinfo{person}{Padmanabhan Pillai}, {and} \bibinfo{person}{Mahadev
  Satyanarayanan}.} \bibinfo{year}{2020}\natexlab{}.
\newblock \showarticletitle{OpenRTiST: End-to-End Benchmarking for Edge
  Computing}.
\newblock \bibinfo{journal}{\emph{IEEE Pervasive Computing}}
  \bibinfo{volume}{19}, \bibinfo{number}{4} (\bibinfo{year}{2020}),
  \bibinfo{pages}{10--18}.
\newblock


\bibitem[\protect\citeauthoryear{{GStreamer Team}}{{GStreamer Team}}{2001}]%
        {gstreamer}
\bibfield{author}{\bibinfo{person}{{GStreamer Team}}.}
  \bibinfo{year}{2001}\natexlab{}.
\newblock \bibinfo{title}{GStreamer: a flexible, fast and multiplatform
  multimedia framework}.
\newblock \bibinfo{howpublished}{\url{https://gstreamer.freedesktop.org/}}.
\newblock
\newblock
\shownote{[Online; accessed 13-June-2021].}


\bibitem[\protect\citeauthoryear{{Holo-Light}}{{Holo-Light}}{2020}]%
        {isar}
\bibfield{author}{\bibinfo{person}{{Holo-Light}}.}
  \bibinfo{year}{2020}\natexlab{}.
\newblock \bibinfo{title}{ISAR SDK – XR Streaming}.
\newblock
  \bibinfo{howpublished}{\url{https://holo-light.com/products/isar-sdk/}}.
\newblock
\newblock
\shownote{[Online; accessed 13-June-2021].}


\bibitem[\protect\citeauthoryear{Hu, Gao, Ha, Wang, Amos, Chen, Pillai, and
  Satyanarayanan}{Hu et~al\mbox{.}}{2016}]%
        {hu2016quantifying}
\bibfield{author}{\bibinfo{person}{Wenlu Hu}, \bibinfo{person}{Ying Gao},
  \bibinfo{person}{Kiryong Ha}, \bibinfo{person}{Junjue Wang},
  \bibinfo{person}{Brandon Amos}, \bibinfo{person}{Zhuo Chen},
  \bibinfo{person}{Padmanabhan Pillai}, {and} \bibinfo{person}{Mahadev
  Satyanarayanan}.} \bibinfo{year}{2016}\natexlab{}.
\newblock \showarticletitle{Quantifying the impact of edge computing on mobile
  applications}. In \bibinfo{booktitle}{\emph{Proceedings of the 7th ACM SIGOPS
  Asia-Pacific Workshop on Systems}}. \bibinfo{pages}{1--8}.
\newblock


\bibitem[\protect\citeauthoryear{Karakaya, Yazici, and Alayyoub}{Karakaya
  et~al\mbox{.}}{2017}]%
        {karakaya2017comparison}
\bibfield{author}{\bibinfo{person}{Ziya Karakaya}, \bibinfo{person}{Ali
  Yazici}, {and} \bibinfo{person}{Mohammed Alayyoub}.}
  \bibinfo{year}{2017}\natexlab{}.
\newblock \showarticletitle{A comparison of stream processing frameworks}. In
  \bibinfo{booktitle}{\emph{2017 International Conference on Computer and
  Applications (ICCA)}}. IEEE, \bibinfo{pages}{1--12}.
\newblock


\bibitem[\protect\citeauthoryear{Karimov, Rabl, Katsifodimos, Samarev,
  Heiskanen, and Markl}{Karimov et~al\mbox{.}}{2018}]%
        {karimov2018benchmarking}
\bibfield{author}{\bibinfo{person}{Jeyhun Karimov}, \bibinfo{person}{Tilmann
  Rabl}, \bibinfo{person}{Asterios Katsifodimos}, \bibinfo{person}{Roman
  Samarev}, \bibinfo{person}{Henri Heiskanen}, {and} \bibinfo{person}{Volker
  Markl}.} \bibinfo{year}{2018}\natexlab{}.
\newblock \showarticletitle{Benchmarking distributed stream data processing
  systems}. In \bibinfo{booktitle}{\emph{2018 IEEE 34th International
  Conference on Data Engineering (ICDE)}}. IEEE, \bibinfo{pages}{1507--1518}.
\newblock


\bibitem[\protect\citeauthoryear{Khare, Sun, Gascon-Samson, Zhang, Gokhale,
  Barve, Bhattacharjee, and Koutsoukos}{Khare et~al\mbox{.}}{2019}]%
        {khare2019linearize}
\bibfield{author}{\bibinfo{person}{Shweta Khare}, \bibinfo{person}{Hongyang
  Sun}, \bibinfo{person}{Julien Gascon-Samson}, \bibinfo{person}{Kaiwen Zhang},
  \bibinfo{person}{Aniruddha Gokhale}, \bibinfo{person}{Yogesh Barve},
  \bibinfo{person}{Anirban Bhattacharjee}, {and} \bibinfo{person}{Xenofon
  Koutsoukos}.} \bibinfo{year}{2019}\natexlab{}.
\newblock \showarticletitle{Linearize, predict and place: minimizing the
  makespan for edge-based stream processing of directed acyclic graphs}. In
  \bibinfo{booktitle}{\emph{Proceedings of the 4th ACM/IEEE Symposium on Edge
  Computing}}. \bibinfo{pages}{1--14}.
\newblock


\bibitem[\protect\citeauthoryear{Lai, Hu, Cui, Sun, Dai, and Lee}{Lai
  et~al\mbox{.}}{2019}]%
        {lai2019furion}
\bibfield{author}{\bibinfo{person}{Zeqi Lai}, \bibinfo{person}{Y~Charlie Hu},
  \bibinfo{person}{Yong Cui}, \bibinfo{person}{Linhui Sun},
  \bibinfo{person}{Ningwei Dai}, {and} \bibinfo{person}{Hung-Sheng Lee}.}
  \bibinfo{year}{2019}\natexlab{}.
\newblock \showarticletitle{Furion: Engineering high-quality immersive virtual
  reality on today's mobile devices}.
\newblock \bibinfo{journal}{\emph{IEEE Transactions on Mobile Computing}}
  \bibinfo{volume}{19}, \bibinfo{number}{7} (\bibinfo{year}{2019}),
  \bibinfo{pages}{1586--1602}.
\newblock


\bibitem[\protect\citeauthoryear{Liu, Li, and Gruteser}{Liu
  et~al\mbox{.}}{2019}]%
        {liu2019edge}
\bibfield{author}{\bibinfo{person}{Luyang Liu}, \bibinfo{person}{Hongyu Li},
  {and} \bibinfo{person}{Marco Gruteser}.} \bibinfo{year}{2019}\natexlab{}.
\newblock \showarticletitle{Edge assisted real-time object detection for mobile
  augmented reality}. In \bibinfo{booktitle}{\emph{The 25th Annual
  International Conference on Mobile Computing and Networking}}.
  \bibinfo{pages}{1--16}.
\newblock


\bibitem[\protect\citeauthoryear{Liu, Zhong, Zhang, Liu, Zhang, Zhang, and
  Gruteser}{Liu et~al\mbox{.}}{2018}]%
        {liu2018cutting}
\bibfield{author}{\bibinfo{person}{Luyang Liu}, \bibinfo{person}{Ruiguang
  Zhong}, \bibinfo{person}{Wuyang Zhang}, \bibinfo{person}{Yunxin Liu},
  \bibinfo{person}{Jiansong Zhang}, \bibinfo{person}{Lintao Zhang}, {and}
  \bibinfo{person}{Marco Gruteser}.} \bibinfo{year}{2018}\natexlab{}.
\newblock \showarticletitle{Cutting the cord: Designing a high-quality
  untethered vr system with low latency remote rendering}. In
  \bibinfo{booktitle}{\emph{Proceedings of the 16th Annual International
  Conference on Mobile Systems, Applications, and Services}}.
  \bibinfo{pages}{68--80}.
\newblock


\bibitem[\protect\citeauthoryear{Liu, Vlachou, Qian, Wang, and Kim}{Liu
  et~al\mbox{.}}{2020}]%
        {liu2020firefly}
\bibfield{author}{\bibinfo{person}{Xing Liu}, \bibinfo{person}{Christina
  Vlachou}, \bibinfo{person}{Feng Qian}, \bibinfo{person}{Chendong Wang}, {and}
  \bibinfo{person}{Kyu-Han Kim}.} \bibinfo{year}{2020}\natexlab{}.
\newblock \showarticletitle{Firefly: Untethered Multi-user {VR} for Commodity
  Mobile Devices}. In \bibinfo{booktitle}{\emph{2020 {USENIX} Annual Technical
  Conference ({USENIX} {ATC} 20)}}. \bibinfo{publisher}{{USENIX} Association},
  \bibinfo{pages}{943--957}.
\newblock
\showISBNx{978-1-939133-14-4}
\urldef\tempurl%
\url{https://www.usenix.org/conference/atc20/presentation/liu-xing}
\showURL{%
\tempurl}


\bibitem[\protect\citeauthoryear{Lopez, Lobato, and Duarte}{Lopez
  et~al\mbox{.}}{2016}]%
        {lopez2016performance}
\bibfield{author}{\bibinfo{person}{Martin~Andreoni Lopez},
  \bibinfo{person}{Antonio Gonzalez~Pastana Lobato}, {and}
  \bibinfo{person}{Otto Carlos~MB Duarte}.} \bibinfo{year}{2016}\natexlab{}.
\newblock \showarticletitle{A performance comparison of open-source stream
  processing platforms}. In \bibinfo{booktitle}{\emph{2016 IEEE Global
  Communications Conference (GLOBECOM)}}. IEEE, \bibinfo{pages}{1--6}.
\newblock


\bibitem[\protect\citeauthoryear{Lu, Wu, Xie, and Hu}{Lu et~al\mbox{.}}{2014}]%
        {lu2014stream}
\bibfield{author}{\bibinfo{person}{Ruirui Lu}, \bibinfo{person}{Gang Wu},
  \bibinfo{person}{Bin Xie}, {and} \bibinfo{person}{Jingtong Hu}.}
  \bibinfo{year}{2014}\natexlab{}.
\newblock \showarticletitle{Stream bench: Towards benchmarking modern
  distributed stream computing frameworks}. In \bibinfo{booktitle}{\emph{2014
  IEEE/ACM 7th International Conference on Utility and Cloud Computing}}. IEEE,
  \bibinfo{pages}{69--78}.
\newblock


\bibitem[\protect\citeauthoryear{Meng, Paul, and Hu}{Meng
  et~al\mbox{.}}{2020}]%
        {meng2020coterie}
\bibfield{author}{\bibinfo{person}{Jiayi Meng}, \bibinfo{person}{Sibendu Paul},
  {and} \bibinfo{person}{Y~Charlie Hu}.} \bibinfo{year}{2020}\natexlab{}.
\newblock \showarticletitle{Coterie: Exploiting frame similarity to enable
  high-quality multiplayer vr on commodity mobile devices}. In
  \bibinfo{booktitle}{\emph{Proceedings of the Twenty-Fifth International
  Conference on Architectural Support for Programming Languages and Operating
  Systems}}. \bibinfo{pages}{923--937}.
\newblock


\bibitem[\protect\citeauthoryear{{Microsoft}}{{Microsoft}}{2019}]%
        {azcustomvision}
\bibfield{author}{\bibinfo{person}{{Microsoft}}.}
  \bibinfo{year}{2019}\natexlab{}.
\newblock \bibinfo{title}{Azure Custom Vision}.
\newblock
  \bibinfo{howpublished}{\url{https://azure.microsoft.com/en-us/services/cognitive-services/custom-vision-service}}.
\newblock
\newblock
\shownote{[Online; accessed 13-June-2021].}


\bibitem[\protect\citeauthoryear{{Microsoft}}{{Microsoft}}{2020}]%
        {azremoterendering}
\bibfield{author}{\bibinfo{person}{{Microsoft}}.}
  \bibinfo{year}{2020}\natexlab{}.
\newblock \bibinfo{title}{Azure Remote Rendering}.
\newblock
  \bibinfo{howpublished}{\url{https://azure.microsoft.com/en-us/services/remote-rendering/}}.
\newblock
\newblock
\shownote{[Online; accessed 13-June-2021].}


\bibitem[\protect\citeauthoryear{{Nvidia Corporation}}{{Nvidia
  Corporation}}{2020}]%
        {cloudxr}
\bibfield{author}{\bibinfo{person}{{Nvidia Corporation}}.}
  \bibinfo{year}{2020}\natexlab{}.
\newblock \bibinfo{title}{NVIDIA CloudXR™ SDK}.
\newblock
  \bibinfo{howpublished}{\url{https://developer.nvidia.com/nvidia-cloudxr-sdk}}.
\newblock
\newblock
\shownote{[Online; accessed 13-June-2021].}


\bibitem[\protect\citeauthoryear{Quigley, Conley, Gerkey, Faust, Foote, Leibs,
  Wheeler, Ng, et~al\mbox{.}}{Quigley et~al\mbox{.}}{2009}]%
        {quigley2009ros}
\bibfield{author}{\bibinfo{person}{Morgan Quigley}, \bibinfo{person}{Ken
  Conley}, \bibinfo{person}{Brian Gerkey}, \bibinfo{person}{Josh Faust},
  \bibinfo{person}{Tully Foote}, \bibinfo{person}{Jeremy Leibs},
  \bibinfo{person}{Rob Wheeler}, \bibinfo{person}{Andrew~Y Ng},
  {et~al\mbox{.}}} \bibinfo{year}{2009}\natexlab{}.
\newblock \showarticletitle{ROS: an open-source Robot Operating System}. In
  \bibinfo{booktitle}{\emph{ICRA workshop on open source software}},
  Vol.~\bibinfo{volume}{3}. Kobe, Japan, \bibinfo{pages}{5}.
\newblock


\bibitem[\protect\citeauthoryear{R{\"o}ger, Bhowmik, and Rothermel}{R{\"o}ger
  et~al\mbox{.}}{2019}]%
        {roger2019combining}
\bibfield{author}{\bibinfo{person}{Henriette R{\"o}ger},
  \bibinfo{person}{Sukanya Bhowmik}, {and} \bibinfo{person}{Kurt Rothermel}.}
  \bibinfo{year}{2019}\natexlab{}.
\newblock \showarticletitle{Combining it all: Cost minimal and low-latency
  stream processing across distributed heterogeneous infrastructures}. In
  \bibinfo{booktitle}{\emph{Proceedings of the 20th International Middleware
  Conference}}. \bibinfo{pages}{255--267}.
\newblock


\bibitem[\protect\citeauthoryear{Schneider, Rambach, and Stricker}{Schneider
  et~al\mbox{.}}{2017}]%
        {schneider2017augmented}
\bibfield{author}{\bibinfo{person}{Michael Schneider}, \bibinfo{person}{Jason
  Rambach}, {and} \bibinfo{person}{Didier Stricker}.}
  \bibinfo{year}{2017}\natexlab{}.
\newblock \showarticletitle{Augmented reality based on edge computing using the
  example of remote live support}. In \bibinfo{booktitle}{\emph{2017 IEEE
  International Conference on Industrial Technology (ICIT)}}. IEEE,
  \bibinfo{pages}{1277--1282}.
\newblock


\bibitem[\protect\citeauthoryear{Taymans, Baker, Wingo, Bultje, and
  Kost}{Taymans et~al\mbox{.}}{2013}]%
        {taymans2013gstreamer}
\bibfield{author}{\bibinfo{person}{Wim Taymans}, \bibinfo{person}{Steve Baker},
  \bibinfo{person}{Andy Wingo}, \bibinfo{person}{Rondald~S Bultje}, {and}
  \bibinfo{person}{Stefan Kost}.} \bibinfo{year}{2013}\natexlab{}.
\newblock \showarticletitle{Gstreamer application development manual (1.2. 3)}.
\newblock \bibinfo{journal}{\emph{Publicado en la Web}} (\bibinfo{year}{2013}).
\newblock


\bibitem[\protect\citeauthoryear{Toshniwal, Taneja, Shukla, Ramasamy, Patel,
  Kulkarni, Jackson, Gade, Fu, Donham, et~al\mbox{.}}{Toshniwal
  et~al\mbox{.}}{2014}]%
        {toshniwal2014storm}
\bibfield{author}{\bibinfo{person}{Ankit Toshniwal}, \bibinfo{person}{Siddarth
  Taneja}, \bibinfo{person}{Amit Shukla}, \bibinfo{person}{Karthik Ramasamy},
  \bibinfo{person}{Jignesh~M Patel}, \bibinfo{person}{Sanjeev Kulkarni},
  \bibinfo{person}{Jason Jackson}, \bibinfo{person}{Krishna Gade},
  \bibinfo{person}{Maosong Fu}, \bibinfo{person}{Jake Donham}, {et~al\mbox{.}}}
  \bibinfo{year}{2014}\natexlab{}.
\newblock \showarticletitle{Storm@ twitter}. In
  \bibinfo{booktitle}{\emph{Proceedings of the 2014 ACM SIGMOD international
  conference on Management of data}}. \bibinfo{pages}{147--156}.
\newblock


\bibitem[\protect\citeauthoryear{{Velimir Mlaker}}{{Velimir Mlaker}}{2018}]%
        {mpipe}
\bibfield{author}{\bibinfo{person}{{Velimir Mlaker}}.}
  \bibinfo{year}{2018}\natexlab{}.
\newblock \bibinfo{title}{MPipe, Multiprocess Pipeline Toolkit for Python}.
\newblock
  \bibinfo{howpublished}{\url{http://vmlaker.github.io/mpipe/index.html}}.
\newblock
\newblock
\shownote{[Online; accessed 13-June-2021].}


\bibitem[\protect\citeauthoryear{{Yu-Ping Wang}}{{Yu-Ping Wang}}{2017}]%
        {rosshm}
\bibfield{author}{\bibinfo{person}{{Yu-Ping Wang}}.}
  \bibinfo{year}{2017}\natexlab{}.
\newblock \bibinfo{title}{shm transport, The shared memory transport package}.
\newblock \bibinfo{howpublished}{\url{http://wiki.ros.org/shm_transport}}.
\newblock
\newblock
\shownote{[Online; accessed 13-June-2021].}


\bibitem[\protect\citeauthoryear{Zaharia, Das, Li, Shenker, and Stoica}{Zaharia
  et~al\mbox{.}}{2012}]%
        {zaharia2012discretized}
\bibfield{author}{\bibinfo{person}{Matei Zaharia}, \bibinfo{person}{Tathagata
  Das}, \bibinfo{person}{Haoyuan Li}, \bibinfo{person}{Scott Shenker}, {and}
  \bibinfo{person}{Ion Stoica}.} \bibinfo{year}{2012}\natexlab{}.
\newblock \showarticletitle{Discretized streams: an efficient and
  fault-tolerant model for stream processing on large clusters}. In
  \bibinfo{booktitle}{\emph{4th $\{$USENIX$\}$ Workshop on Hot Topics in Cloud
  Computing (HotCloud 12)}}.
\newblock


\bibitem[\protect\citeauthoryear{Zhang, Sun, Shea, Liu, and Zhang}{Zhang
  et~al\mbox{.}}{2019}]%
        {zhang2019rendering}
\bibfield{author}{\bibinfo{person}{Lei Zhang}, \bibinfo{person}{Andy Sun},
  \bibinfo{person}{Ryan Shea}, \bibinfo{person}{Jiangchuan Liu}, {and}
  \bibinfo{person}{Miao Zhang}.} \bibinfo{year}{2019}\natexlab{}.
\newblock \showarticletitle{Rendering multi-party mobile augmented reality from
  edge}. In \bibinfo{booktitle}{\emph{Proceedings of the 29th ACM Workshop on
  Network and Operating Systems Support for Digital Audio and Video}}.
  \bibinfo{pages}{67--72}.
\newblock


\end{thebibliography}
